# Structures of Neutrino Mass Spectrum and Lepton Mixing: Results of Violated Mirror Symmetry

**Igor T. Dyatlov ***
**Scientific Research Center "Kurchatov Institute"**
**Petersburg Institute of Nuclear Physics, Gatchina, Moscow, Russia**

The specific violated mirror symmetry model is capable of generating the observed lepton weak mixing matrix with a structure similar to the observed one that almost lacks any visible regularities (the "flavor riddle"). The peculiarities of the Standard Model (SM): quark and lepton mass hierarchy and the neutrino spectrum different from this hierarchy appear to be necessary conditions for reproduction of such a structure. The inverse order of the neutrino spectrum and a small value of the mass $m_3$ are here two other necessary conditions. The smallness of the angle $\theta_{13}$ is determined then just by small mass ratios in the hierarchical lepton spectrum. The explanation is proposed for differences between the neutrino spectrum and other fermion spectra of SM. The inverse order of the neutrino spectrum and the observed $\theta_{13}$ angle permit evaluation of the absolute values of neutrino masses: $m_1 \sim m_2 \sim 0.05$ eV, $m_3 \lesssim 0.01$ eV.

## 1. Introduction

The observed neutrino mass spectrum, with two states "1,2" very close in mass and one state "3" remote from the first two, differs from spectra of all other SM fermions, i.e., quarks and charged leptons, which obey a strict hierarchical structure. The lepton weak mixing matrix (WMM) is also completely different from the quark WMM—that is, the Cabibbo-Kobayashi-Maskawa (*CKM*) matrix [1].

In this paper, these differences are explained by means of a model [2,3] in which SM fermion masses are formed as a result of spontaneous violation of initial mirror symmetry, i.e., in SM supplemented with heavy analogs with opposite (left-right) weak properties [4]. At that, the notion of "mirror symmetry" (MS) differs in [2,3] and [4].

For the observed quark mass hierarchy, such a system reproduces all qualitative features of the *CKM* matrix. The masses of mirror fermions must be large for such a reproduction [5].

For leptons, the mirror model produces factors that are indicative of the exceptional smallness of SM neutrino masses, and requires the inverse order of these masses (the two states close to each other are the heavier) and the Dirac nature of both mirror and "normal" neutrinos [2].

The failure of the neutrino mass spectrum to obey simple hierarchy results here from two independent spectra of neutral mirror masses—Dirac and Majorana—being jointly involved in the formation of mass matrices. If taken separately, however, both of these spectra have masses arranged into the "normal" hierarchical order, similar to spectra of all other SM fermions.

* E-mail: dyatlov@thd.pnpi.spb.ru

Note that the mechanism of neutrino mass formation in this case is essentially different from the known see-saw scenario [6]; neutrinos remain Dirac despite the presence of specific type Majorana terms in the structure of their spectrum.

Thus, mass hierarchy is a common feature of all spectra involved in mass formation for all fermions, including neutrinos. We are dealing here with an as yet unexplained universal dynamical mechanism (see [7] for attempts to explain mass hierarchy).

The mirror scenario also provides explanation for a different form of the lepton WMM. The observed neutrino mass spectrum, its inverse order, and charged lepton mass hierarchy simultaneously produce the Pontecorvo–Maki–Nakagawa–Sakata (*PMNS*) WMM [1] with properties fundamentally different from the properties of the *CKM* matrix. In addition, the mass of the lightest (in inverse spectrum) neutrino must be small: $m_3 \ll m_1 \sim m_2$. All experimentally determined properties of the *PMNS* matrix are reproduced qualitatively.

The smallness of the Daya-Bay angle [8] $\theta_{13}$—the main peculiarity of the *PMNS* matrix—results from the orthogonality, in generation space, of the electron and remote neutrino "3" wavefunctions; this orthogonality being produced, neglecting small mass ratios, by the mirror model. The value $|\sin\theta_{13}|$ is defined precisely by these small mass ratios, particularly, it seems, by neutrino mass ratios $(m_3/m_1) \sim (m_3/m_2)$. Other mass ratios appear to make lesser contributions to $\sin\theta_{13}$.

The inverse order of the spectrum and the required smallness of the mass $m_3$ also permit here estimation of the absolute values of neutrino masses. The observed value $|\sin\theta_{13}| \sim 0.14 - 0.16$ corresponds to the unknown mass of the lightest neutrino "3": $m_3 \lesssim 0.01$ eV. Note that for the inverse order, large masses are defined then by the differences $\Delta m_{13}^2 \approx \Delta m_{23}^2$ [1] and equal $m_1 \approx m_2 \approx 0.05$ eV. No additional constraints on model constants (besides those that are imposed by mass hierarchy) are necessary for the reproduction of the qualitative picture.

All results appear to occur only for the inverse order of SM neutrino mass spectrum ($\Delta m_{13}^2 \approx \Delta m_{23}^2 > 0$). Establishing the order of this spectrum would be the crucial step to test the mirror hypothesis [9]. The Dirac nature of neutrino required by the scenario makes double beta-decay impossible. New mirror particles may be too heavy.

Our investigation is by no means a rigorous logical theory of lepton WMM generation. It is rather an attempt to find, with the help of the theory and observed phenomenology, conditions and mechanisms capable of producing as yet unexplained properties of the *PMNS* lepton matrix (see [10] for bibliography on this problem).The system with spontaneous breaking of mirror symmetry possesses required, for this purpose, qualities.

This paper is the continuation of other articles by the author [2,3,5] and is largely based on results from these articles. Section 2 contains the main concepts from the previous articles, as well as reproducing some of the formulae required for the purpose of this paper. However, the largest portion of scenario descriptions and calculations, although necessary for better understanding of the presented material, remains outside the scope of this paper.

Section 2 formulates the MS scenario used in this paper and discusses conditions for MS breaking and its implications. In Section 3, neutrino mass spectrum and its observed characteristics [1] are explained within the framework of the mirror hypothesis. Section 4 shows how MS breaking allows, using observed SM properties, reproduction of the qualitative features of the lepton WMM—that is, the *PMNS* matrix. Section 5 Conclusion summarizes the results of this study and outlines further questions raised by the mirror symmetry scenario under consideration.

## 2. Mirror Symmetry and Mirror Symmetry Violation

The common understanding of mirror systems [4] consists in adding to the existing fermion SM multiplets analogs with different masses and different weak properties—that is, replacing left-handed (*L*) chiral currents by right-handed (*R*) ones. The expansion of SM then pursues only one goal: the elimination of the paradox resulting from direct parity non-conservation, i.e., creating the possibility to determine physically the left- or right-handed character of the coordinate system being used.

In [2,3], no intentional introduction of "mirror" multiplets is carried out. The differences in "mirror" properties are consequences of the system dynamics. We consider usual massive Dirac fermions (i.e., $R + L$ states) with weak $SU(2)$-isospins $T_W = 1/2,0$:

$$\Psi_{LR} = \psi_L + \Psi_R, \quad T_W = \frac{1}{2}; \quad \Psi_{RL} = \psi_R + \Psi_L, \quad T_W = 0; \tag{1}$$

Three generations with indices $a = 1,2,3$ and two flavors up($\bar{u}$) and down($\bar{d}$) are assumed. The vector currents of $\Psi_{LR}$- particles with $T_W \neq 0$ determine interactions with the weak boson $W$.

The obvious MS:

$$L \leftrightarrow R, \quad \psi \leftrightarrow \Psi, \tag{2}$$

is violated by spontaneous generation of additional heavy masses for only the $\Psi_L,\Psi_R$-components. This results in separation of (1) into

standard multiplets $\left\{ \psi_L(T_W = \frac{1}{2}), \, \psi_R(T_W = 0) \right\}$

and mirror multiplets $\left\{ \Psi_R(T_W = \frac{1}{2}), \, \Psi_L(T_W = 0) \right\}$.

The initial MS system in which all fermions are expressed only in terms of the operators $\Psi_{LR}$ and $\Psi_{RL}$ is described by a sum of SM Lagrangians for $\psi$ and $\Psi$. The sum includes gauge and Yukawa interactions. The latter are expressed in terms of products consisting of operators (1):

$$h \, \bar{\Psi}_{LR}\Psi_{RL} = h\big(\bar{\psi}_L\psi_R + \bar{\Psi}_R\psi_L\big); \quad h^+\bar{\Psi}_{RL}\Psi_{LR} = h^+\big(\bar{\psi}_R\psi_L + \bar{\Psi}_L\Psi_R\big). \tag{3}$$

Similar to SM, the constants $h$, in general terms, represent arbitrary matrices of the generation indices $a, b = 1,2,3$, which differ for the $f = \bar{u}, \bar{d}$ flavors.

Only MS masses of the Dirac fermions $\Psi_{LR}$ and $\Psi_{RL}$ are not sums of $\psi$ and $\Psi$ contributions, while they carry out transitions between these components:

$$\mathcal{L}_1 = A\bar{\Psi}_{LR}\Psi_{LR} + B\bar{\Psi}_{RL}\Psi_{RL} = A\bar{\psi}_L\Psi_R + B\bar{\psi}_R\Psi_L + c.c. \tag{4}$$

The masses $A$ and $B$ can naturally be considered as diagonal real matrices. Eq.(4) should be written out for both flavors $\bar{u}$ and $\bar{d}$. The weak $SU$(2)-symmetry requires that the masses $A$ for $\bar{u}$ and $\bar{d}$ be equal, while the masses $B$ may differ from each other:

$$A^{(\bar{u})} = A^{(\bar{d})} \equiv A, \quad B^{(\bar{u})} \neq B^{(\bar{d})}. \tag{5}$$

Possible models of spontaneous symmetry breaking are discussed in [3,11]. In these models, Yukawa couplings are chosen such that they generate mass terms for $\Psi$-states only:

$$\mathcal{L}_2 = \mu\bar{\Psi}_R\Psi_L + \mu^+\bar{\Psi}_L\Psi_R. \tag{6}$$

where $\mu$, similar to SM, are matrices of generation indices $\mu^{(\bar{u})} \neq \mu^{(\bar{d})}$. For separation of the $\psi$- and $\Psi$-components in the Yukawa couplings, two different bosons must be used, scalar and pseudoscalar. Certain of their combinations result in vacuum averages $\eta$, which transfer the MS-system into one of two mirror "worlds", either light $\psi$ and heavy $\Psi$ or light $\Psi$ and heavy $\psi$. It is obvious that

$$\mu = h\,\eta. \tag{7}$$

In both cases, all properties except the weak ones, $R{\leftrightarrow}L$, must be identical. Otherwise, the main mirror principle would be violated, making it impossible to determine physically the *L,R*-character of the coordinate system. Generally speaking, if this identity were lacking, it would be possible to determine which of the two states the system is in and thus fixate *L* or *R.*

The operators $\Psi_{LR}$ and $\Psi_{RL}$ can be transformed in generation space with unitary matrices $U_{LR}$ and $U_{RL}$. As with SM, the Yukawa couplings $h$, and, consequently, mass matrices (7), $\mu$, should preferably be made diagonal. In this case, $A$ and $B$ in Eq.(4) become Hermitian matrices. Reproduction of quark WMM properties requires [5] that condition (5) for $A$ be fulfilled not only for the diagonal masses (4), but also for all elements of the transformed matrices. This means that the $U_{LR}$ matrices that diagonalize (3) and (6), and which transform the isospinors $\Psi_{LR}$, must be independent of the flavor:

$$U_{LR}^{(\bar{u})} = U_{LR}^{(\bar{d})} \equiv U_{LR}, \quad A = U_{LR}A_{diag}U_{LR}^+. \tag{8}$$

Then, weak interactions and all other parts of the Lagrangian (except (4)) will remain diagonal upon diagonalization of the Yukawa couplings. Unlike SM, this operation does not produce a WMM; all weak currents remain diagonal. The WMM formation mechanism is different here.

Conditions (5) and (8) can be considered as direct consequences of $SU(2)$-symmetry in the MS theory. Their role consists in restricting forms of the possible Yukawa matrices $h$ and mass matrices $\mu$.

In terms of diagonal $\mu$, the mass matrix of SM fermions $\psi$, defined by Eqs.(4) and (6), has the separable form [3]:

$$\left(M_{RL}^{(f)}\right)_a^b \simeq \sum_{n=0}^{2} A_a^n \frac{1}{\mu_n^{(f)}} B_n^{(f)+b},$$

(9)

where $n = 0,1,2$ is the numeration of diagonal states (6), chosen based on the order of the hierarchy (see (12)).

Eq.(9) is an approximate solution of the problem. It takes into account contributions of the pole diagrams, Fig.1, provided that the masses $m_i$ of SM generations calculated from Eq.(9) are significantly lighter than the masses $m_n$ of mirror fermions. At that, the momenta $|\hat{p}| = m_i \ll \mu$ in the propagators in Fig.1 can be neglected—that is:

$$A \frac{1}{\mu} B \ll \mu.$$

(10)

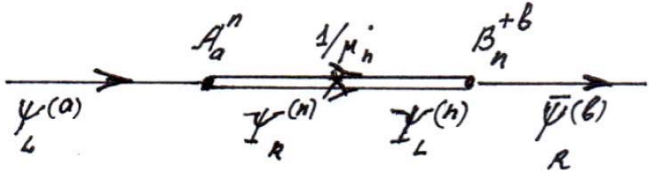

Fig.1. Fermion mass matrix formation in MS violated models

The observed masses of SM fermion generations (except neutrinos) obey simple hierarchy:

$$m_I \ll m_{II} \ll m_{III}.$$

(11)

We assume that this property of the mass matrix (9) is related with the hierarchical order of the mirror mass spectrum:

$$\mu_2 \gg \mu_1 \gg \mu_0.$$

(12)

Let us assume that it is the denominators (12) that define the value of the terms in Eq.(9). This leads to a remarkable structural uniformity of spectra of all masses violating MS (see Section 3), which allows us to propose a common mechanism of their origin. Numbers $n = 0,1,2$ in (9) and (12) are the orders of hierarchy repeating the spectrum (11).

The phenomenological conditions (11) and (12) are fulfilled only for quark and charged lepton generations; Eq.(9) is also used for them.

For any complex matrices $A$ and $B^{(f)}$ (provided they do not violate the hierarchy of the terms in (9)), Eq.(9) results in the mass spectrum (11) of SM generations. Simultaneously, with the matrix conditions (8) fulfilled, all qualitative properties of quark WMMs—the *CKM* matrix—are reproduced. This means reproduction of the hierarchy of elements parametrized in the Wolfenstein matrix [1,12]. A solution for these problems is presented in [5].[1]

This solution has an important, for further discussion, property in the lowest hierarchical order: the left eigenfunctions of the separable matrices (9) are only dependent on the matrices $A$—that is, on the generation space vectors $A_0^a$, $A_1^a$ and $A_2^a$. For hierarchy (11), we obtain zero-approximation *L*-wavefunctions:

$$\Phi_{III}^{(0)} = \frac{1}{N_{III}} A_0 \,, \quad \Phi_{II}^{(0)} = \frac{1}{N_{II}} \Big( A_1 - \frac{(A_0^+, A_1)}{|A_0|^2} A_0 \Big), \quad \Phi_I^{(0)} = \frac{1}{N_I} [A_0^+, A_1^+], \tag{13}$$

where [...] is a vector product. For the unitary WMM, the orthonormality of the functions (13), with $A$ independent of flavor, means that

$$V_{IK} = \Big( \Phi_I^{(d)+}, \, \Phi_K^{(\bar{u})} \Big) \tag{14}$$

has diagonal elements close to unity and non-diagonal elements small in terms of mass ratio orders (12) that are now also dependent on $B^{(f)}$. This is how the *CKM* matrix structure forms [5]. Eqs.(9) and (13) will be used in further discussion in Section 4.

The independence of the isodoublet matrix $A$ from flavor is the major condition for the mirror approach to be consistent with the observed characteristics. This independence must be present in the MS-Lagrangian for any rotation of generation space (8) prior to MS violation. This is a consequence of the *SU*(2)-symmetry, which is also supported by the observed phenomenology of the quark WMM and, as shown later in this paper, by the possibility to reproduce the properties of the lepton WMM. The dependence of $A$ on flavor would transform the quark WMM into an arbitrary unitary matrix without the Wolfenstein hierarchical structure [12].

The requirement of the independence of $A$ from flavor restricts the arbitrariness of Yukawa coupling matrices in mirror models, making them different from the absolutely general form of the Yukawa couplings in SM.

Note, however, that selection of concrete properties of the Yukawa matrices $h$ is essential only when considering neutrino. It is sufficient to have condition (12) and the independence of the arbitrary complex matrix $A$ from flavor in order to reproduce the mass hierarchy and quark WMM with the matrix (9).

---

[1] Eq.(36) in [5] has a typo: the mass ratio $(m_c/m_t)$, which should be $\approx 0.008$, is erroneously written as 0.08.

## 3. Neutrino Mass Spectrum

The identity of quark and charged lepton spectrum hierarchies allows us again to use the mechanism described in the previous section (see Fig.1) for charged leptons. The lepton factors $A$, $B$, $h$, and $\mu$ possess analogous qualities that were chosen for quarks. The mass matrix and eigenfunctions of charged leptons are expressed in terms of the lepton parameters again with the use of Eqs.(9) and (13).

The unique properties of the neutrino system indicate that neutrino masses are formed by means of a different mechanism, which becomes possible through the use of the Majorana analogs of Yukawa couplings acceptable for neutral fermions. For the MS case, we again write them out only in terms of the operators (1) [2].

Let us first mention an important condition that, in our mirror approach, restricts the form of Yukawa and Majorana coupling matrices. Here, parity non-conservation $P$ must occur only through weak interactions, when MS violation results in different masses of $L$- and $R$-weak current particles. Parity non-conservation through other MS interactions contradicts the very idea of impossibility to physically distinguish between the $L$- and $R$-coordinate systems.

For the observed properties of the quark system, parity conservation changes the number of $h$- and $\mu$-parameters but not the spectrum structure and WMM derived from Eq.(3). For the observed properties of leptons, parity conservation in the initial MS-Lagrangian becomes a necessary prerequisite for the reproduction of observed qualities.

This means that the MS-Lagrangian permits only Hermitian matrices of Yukawa couplings and mass matrices $\mu$ (7):

$$\mu \bar{\Psi}'_R \Psi'_L = \mu \bar{\Psi}_L \Psi_R \equiv \mu^+ \bar{\Psi}_L \Psi_R \,; \quad \mathcal{P} : \Psi' = e^{i\alpha}\gamma_0 \Psi, \quad L \leftrightarrow R, \quad \mu = \mu^+. \tag{15}$$

The Hermitian character of $h$ and $\mu$ means that, in MS, Yukawa couplings can be diagonalized by one unitary matrix, i.e., one can assume that $U_{LR} = U_{RL} = U$—all matrices are independent of flavor. For Majorana couplings producing the Majorana mass terms [3]:

$$\mathcal{M}_L \Psi_L^T C \Psi_L + \mathcal{M}_R \Psi_R^T C \Psi_R + c.c. \,, \tag{16}$$

only those scenarios that conserve parity: i.e., $\mathcal{M}_L = \pm \mathcal{M}_R$, are acceptable:

$$\Psi_L^T C \Psi'_L = \pm \Psi_R^T C \Psi_R \,; \quad \mathcal{P} : \Psi' = i\gamma_0 \Psi, \quad \text{и} \quad \Psi' = \gamma_0 \Psi, \quad L \leftrightarrow R \,. \tag{17}$$

The only scenario that matches the observed phenomenology is [2,3]:

$$\mathcal{M}_L = -\mathcal{M}_R \equiv \mathcal{M}. \tag{18}$$

Choosing the (+) sign results in the Majorana character of neutrino, in the mass formulae this variant can easily be matched with a WMM, such as the *CKM*, but that are completely deprived of qualities promoting WMM properties of real SM neutrinos.

Choosing (18) results in the Dirac character of mirror and standard neutrinos. Both components $\Psi_R$ and $\Psi_L$ are involved in the formation of single Dirac mirror states $n = 0,1,2$, which become formed as pairs of Majorana masses with equal absolute mass values. The combination of $\Psi_R$ and $\Psi_L$ into the same particle means that there should be a possibility to simultaneously reduce the matrices $\mu$ and $\mathcal{M}$ to a diagonal form, in other words, the equality of the diagonalizing matrices $U$.

The Dirac nature of SM neutrinos $(\psi_R, \psi_L)$ is supported by direct calculations of the neutrino mass matrix using Eqs.(8), (9) and (15) [2]. The Majorana part of this matrix is equal to zero. The Dirac part of the mass matrix has the form:

$$(\mathcal{M}_{RL}^{(\nu)})_a^b \approx \sum_{n=0}^{2} A_a^n \Big(\frac{\mu_n}{\mathcal{M}_n^2}\Big)^{(\nu)} B_n^{(\nu)+b} = \sum_{n=0}^{2} A_a^n \frac{1}{\mu_n^{(\nu)}} B_n^{(\nu)+b} \Big(\frac{\mu_n}{\mathcal{M}_n}\Big)^{2(\nu)}. \tag{19}$$

At $\mathcal{M} \gg \mu$, Eq.(19) provides explanation for the exceptional smallness of neutrino masses compared to charged lepton masses:

$$m_\nu \sim \Big[ A \frac{1}{\mu} B^+ \Big(\frac{\mu}{\mathcal{M}}\Big)^2 \Big]^{(\nu)} \sim m_{ch} \Big(\frac{\mu}{\mathcal{M}}\Big)^{2(\nu)}. \tag{20}$$

where $m_{ch}$ are masses of the order of charged leptons. Differences from Eq.(8) for $\bar{u}$- and $\bar{d}$-quarks are obvious.

One should assume that the hierarchy of Dirac mass parameters for mirror neutrinos repeats qualitatively the mass hierarchy of mirror charged leptons that is inverse to the hierarchy of SM leptons and consistent with spectrum (12). This relationship is also supported by the required equality of the diagonalizing matrices (8). This same identity exists for the Dirac masses of $\bar{u}$- and $\bar{d}$-quark families. In Eq.(19), however, the masses $\mu^{(\nu)}$ appear in the numerators, which permits one to propose an inverse order ($0 \leftrightarrow 2$) for the terms of summation over $n$ and for SM neutrino masses. In the MS approach, this hypothesis leads to, and is supported by, important phenomenological implications. Therefore, we assume that the inverse order of SM neutrino masses, as prompted by Eq.(19), does exist. Then, the lepton WMM will be completely dissimilar to the quark WMM. Its form and correspondence with the observed *PMNS* matrix are built and discussed in the following section.

The observed neutrino spectrum, however, does not demonstrate simple hierarchical behavior. For the inverse mass order, there are two "heavy" states "1,2" close to each other and a light state "3", which is considered as hierarchically remote from the first two: $m_3 \ll m_1 \sim m_2$. Only in this exceptional case, the formulae in Section 4 leading to the agreement with the *PMNS* matrix properties will be correct. The hierarchy of the charged mirror masses $\mu_1$ and $\mu_2$ is inverse relative to the mass ratio of SM light leptons $m_e/m_\mu$. For our calculation, we assume that the ratio $(\mu_1/\mu_2)^{(\nu)}$ has the same order, i.e.:

$$\frac{\mu_1}{\mu_2} \sim \frac{m_\mu}{m_e} \sim 10^2. \tag{21}$$

This corresponds to the mass ratio of charged leptons produced by Eq.(9). Then, from (20) we have:

$$\left(\frac{\mathcal{M}_1}{\mathcal{M}_2}\right)^2 \sim \frac{m_e}{m_\mu}\left(\frac{\mu_1}{\mu_2}\right)^2 \sim \frac{m_\mu}{m_e} \sim 10^2. \tag{22}$$

The estimate (22) shows that the matrix (19) and the observed neutrino spectrum require that unknown Majorana parameters $\mathcal{M}$ also obey spectrum hierarchy. This means that spectra of all MS-breaking mass parameters $\mu, \mathcal{M}$ obey the same principle. Hierarchy becomes a universal property of mass spectra. The unique features of the observed neutrino spectrum result from the mutual action of the two spectra with a strictly hierarchical mass order.

This universality is deemed to be evidence of the dynamic nature of spontaneous MS breaking, the formation of the Yukawa and Majorana couplings. In our discussion, unknown dynamics are modelled by introducing fundamental scalars similar to the Higgs bosons of SM.

One can compare Eq.(20) and its consequence (22) with the result that would be generated by the see-saw formula $\sim (\mu^2/\mathcal{M})$ for neutrino masses, again for inverse order. The ratio of the parameters $\mathcal{M}$ would then significantly exceed any known ratios of the generation hierarchy $(\mathcal{M}_1/\mathcal{M}_2) \gtrsim 10^4$.

To conclude this section, let us note that hierarchy as a universal property of spectra takes place only if neutrino masses obey the inverse order and provides additional support to such order and the entire MS approach.

## 4. Neutrino Mixing

Let us first consider inverse neutrino spectrum with simple mass hierarchy ($\nu$ is omitted):

$$m_1 \gg m_2 \gg m_3. \tag{23}$$

The hierarchical smallness of $m_3$ appears to be a necessary condition for the correspondence with lepton WMM properties. Further consideration of the observed inverse spectrum

$$m_1 \approx m_2 \gg m_3 \tag{24}$$

requires first solving the problem with hierarchy of Eq.(23) [2].

The orthonormalized left eigenfunctions of the separable matrix (19) in the lowest approximation of the hierarchy again depend on only vectors $A_n$ and coincide with the expressions (13) at $A_0 \leftrightarrow A_2$. For the three neutrino states, 1, 2, 3, we have:

$$\Phi_1^{(0)} = \frac{1}{N_1} A_2 \,, \quad \Phi_2^{(0)} = \frac{1}{N_2} \Big( A_1 - \frac{(A_2^+ A_1)}{|A_2|^2} A_2 \Big), \quad \Phi_3^{(0)} = \frac{1}{N_3} \big[ A_2^+, A_1^+ \big]. \tag{25}$$

The normal hierarchy (11) of charged particles preserves eigenfunctions (13) for such leptons: $III \rightarrow \tau$, $II \rightarrow \mu$, $I \rightarrow e$, with the same flavor-independent vectors $A$ used in Eq.(25). The scalar products (14) for the functions (13) and (25) result, for leptons (23), in a WMM completely different from the *CKM* matrix. The new matrix lacks diagonal unities and non-diagonal element hierarchy (see Eq.(41) in [2]).

Thus, the problem of the lepton WMM in the MS approach can be solved based on the characteristics of charged leptons only, with no consideration for the Yukawa or Majorana properties of neutrinos themselves: vectors $A$ do not depend on $\nu, \ell$-flavor. (In Section 5, we will discuss possible properties of those characteristics of neutrino systems that are not used in our scenario.)

The charged lepton matrix $\mu^{(\ell)}$ (further on, $\mu$) must be Hermitian (parity conservation) and correspond to the inverse hierarchy (12) of mirror analogs of SM leptons $e, \mu, \tau$. In the real lepton WMM [1], the contribution of CP-violating complexities does not affect significantly the main structure of the matrix. Therefore, for simplicity and to clarify our interpretation ($A$ are real three-dimensional vectors), let us consider the real symmetric matrix $\mu$.

Such a matrix, with a hierarchy of eigenvalues, can be built by extending the known see-saw mechanism [6] to the system of three states. We have:

$$\mu = \begin{vmatrix} M & m_1 & m_2 \\ m_1 & m & 0 \\ m_2 & 0 & 0 \end{vmatrix} \sim \begin{vmatrix} M & m_2 & m_1 \\ m_2 & 0 & 0 \\ m_1 & 0 & m \end{vmatrix}, \quad M \gg m_i \,. \tag{26}$$

Matrix (26) is chosen such that it has only one large element present (the energy scale). Later on, it will be clear that this is the only important and necessary feature that ensures the appearance of lepton WMM properties. By changing the generation indices in $\psi_{LR}$ and $\psi_{RL}$, this element can be placed in position (1,1). Other forms of the matrix $\mu$ (or $h$) with a hierarchy of eigenvalues and the reasons why we consider them unsuitable for our scenario are discussed in Section 5.

In (26), the elements equal to zero could mean negligible quantities that are too small compared to those taken into consideration. Their choice is also influenced by another condition of the hierarchy prompted by the see-saw mechanism: the determinant does not contain the large scale $M$. The matrix (26) leads to the characteristic equation:

$$(-\mu)^3 + (-\mu)^2 (M + m) + (-\mu)(Mm - m_{12}^2) - m\, m_2^2 \;=\; 0, \quad m_{12}^2 = m_1^2 + m_2^2 \,. \tag{27}$$

As known, its coefficients are expressed through the roots $\mu_i$ and equal, respectively,

$$\mu_2 + \mu_1 + \mu_0, \quad \mu_2(\mu_1 + \mu_0) + \mu_1\mu_0, \quad \mu_2\mu_1\mu_0, \quad \mu_2 \gg \mu_1 \gg \mu_0. \tag{28}$$

At large $M$, the roots are easily found from (27) and (28) to any precision. For the hierarchy of eigenvalues, their order of magnitude is defined by ratios of the next, consecutive coefficients (27). Root systems depend on ratios between the terms in the coefficients at $(-\mu)$:

$$Mm > m_{12}^2 \quad \text{(a)}, \quad Mm < m_{12}^2 \quad \text{(b)}. \tag{29}$$

The formulae that follow are written out for the case when the inequalities (29) mean "much greater" or "much less". While not changing the essence of the process of formation of *PMNS* matrix properties, simplification of the formulae facilitates the understanding of the MS-mechanism being discussed.

With an accuracy to $m_i/M$, we obtain for both cases in (29):

$$
\begin{aligned}
(a): \quad &\mu_2 = M + m_{12}^2/M & (b): \quad &\mu_2 = M + m_{12}^2/M \\
&\mu_1 = m - m_1^2/M & &\mu_1 = -m_{12}^2/M + m(m_1^2/m_{12}^2) \\
&\mu_0 = -m_2^2/M & &\mu_2 = (m_2^2/m_{12}^2)\, m
\end{aligned}
\tag{30}
$$

Note that, for (30b), Eq.(29) entails:

$$(b): \quad m = \frac{\tilde{m}^2}{M} < \frac{m_{12}^2}{M}, \tag{31}$$

so that in our formulae for case (b) $m \ll (m_{12}^2/M)$. The sign of the mass can mean a different parity of the fermion. Similar to the see-saw mechanism [6], this does not have any implications for our scenario. Dissimilarities between the two cases, (a) and (b), manifest themselves at $M \to \infty$. In case (a), in addition to $M \to \infty$, one more finite term $m$ remains in the formulae. In case (b), all masses with the exception of $M \to \infty$ vanish. Here, case (b) completely corresponds to the see-saw mechanism.

By finding eigenfunctions for each of the roots in cases (a) and (b), we can build the orthogonal matrices U that diagonalize (26) (again with an accuracy to $(m_i/M)^2$):

$$
(a): \quad U = \begin{vmatrix} 1 & \frac{m_1}{M} & \frac{m_2}{M} \\ \frac{m_1}{M} & -1 & -\frac{m_1 m_1}{mM} \\ \frac{m_2}{M} & \frac{m_1 m_2}{mM} & -1 \end{vmatrix}; \quad (b): \quad U = \begin{vmatrix} 1 & \frac{m_{12}}{M} & \frac{m m_1 m_2}{m_{12}^3} \\ \frac{m_2}{M} & -\frac{m_2}{m_{12}} & \frac{m_1}{m_{12}} \\ \frac{m_1}{M} & -\frac{m_1}{m_{12}} & -\frac{m_2}{m_{12}} \end{vmatrix}. \tag{32}
$$

The diagonal mass matrix of isodoublets, $\tilde{A} = (\tilde{A}_1, \tilde{A}_2, \tilde{A}_3)$ (4), can be transformed using the matrix $U$: $U\, \tilde{A} U^+$. For the vector-columns $A_1, A_2, A_3$ in (9) and (19), we have for the lepton matrices $\ell$ and $\nu$:

(a):

$$
A = \begin{array}{c|ccc}
 & A_2 & A_1 & A_0 \\
\hline
 & \tilde{A}_1 & \left(\tilde{A}_1 - \tilde{A}_2\right)m_1/M & \left(\tilde{A}_1 - \tilde{A}_3\right)m_2/M \\
 & \left(\tilde{A}_1 - \tilde{A}_2\right)m_1/M & \tilde{A}_2 & \left(\tilde{A}_3 - \tilde{A}_2\right)\dfrac{m_1 m_2}{mM} \\
 & \left(\tilde{A}_1 - \tilde{A}_3\right)m_2/M & \left(\tilde{A}_3 - \tilde{A}_2\right)\dfrac{m_1 m_2}{mM} & \tilde{A}_3
\end{array}; \tag{33}
$$

$$
A = \begin{array}{c|ccc}
 & A_2 & A_1 & A_0 \\
\hline
 & \tilde{A}_1 & \left(\tilde{A}_1 - \tilde{A}_2\right)\dfrac{m_2}{M} + \tilde{A}_3\dfrac{m_1^2 m_2 m}{m_{12}^3} & \left(\tilde{A}_1 - \tilde{A}_2\right)\dfrac{m_1}{M} - \tilde{A}_3\dfrac{m_2^2 m_1 m}{m_{12}^3} \\
 & \left(\tilde{A}_1 - \tilde{A}_2\right)\dfrac{m_2}{M} + \tilde{A}_3\dfrac{m_1^2 m_2 m}{m_{12}^3} & \tilde{A}_3\dfrac{m_1^2}{m_{12}^2} + \tilde{A}_2\dfrac{m_2^2}{m_{12}^2} & \left(\tilde{A}_2 - \tilde{A}_3\right)\dfrac{m_1 m_2}{m_{12}^2} \\
 & \left(\tilde{A}_1 - \tilde{A}_2\right)\dfrac{m_1}{M} - \tilde{A}_3\dfrac{m_2^2 m_1 m}{m_{12}^3} & \left(\tilde{A}_2 - \tilde{A}_3\right)\dfrac{m_1 m_2}{m_{12}^2} & \tilde{A}_3\dfrac{m_2^2}{m_{12}^2} + \tilde{A}_2\dfrac{m_1^2}{m_{12}^2}
\end{array} \tag{34}
$$

In the lowest approximation of the mass hierarchy, we obtain a diagonal matrix for case (33)(a), i.e., vectors $A_n$ orthogonal in generation space. For inverse neutrino mass hierarchy, the lepton WMM (14) appears to be a unitary diagonal matrix, and as such is not suitable for *PMNS* matrix generation.[2]

In the lowest approximation of the mass hierarchy, we obtain for case (34)(b):

$$
A = \begin{array}{c|ccc}
 & A_2 & A_1 & A_0 \\
\hline
 & \tilde{A}_1 & 0 & 0 \\
 & 0 & \tilde{A}_3\dfrac{m_1^2}{m_{12}^2} + \tilde{A}_2\dfrac{m_2^2}{m_{12}^2} & \left(\tilde{A}_2 - \tilde{A}_3\right)\dfrac{m_1 m_2}{m_{12}^2} \\
 & 0 & \left(\tilde{A}_2 - \tilde{A}_3\right)\dfrac{m_1 m_2}{m_{12}^2} & \tilde{A}_3\dfrac{m_2^2}{m_{12}^2} + \tilde{A}_2\dfrac{m_1^2}{m_{12}^2}
\end{array} \tag{35}
$$

Fig.2 shows the $A_n$ vectors from Eq.(35) and orthonormalized vectors (13) and (25), which, in the approximation being considered, are wavefunctions of SM particles. Note that the directions are matched:

$$
\begin{aligned}
&\text{for } \nu: & &\nu_1 \sim A_2, \ \nu_2 \sim A_1 - \cos\alpha_{12}A_2, \ \nu_3 \sim [A_1, A_2]; \\
&\text{for } \ell^{\pm}: & &\tau \sim A_0, \ \mu \sim A_1 - \cos\alpha_{01}A_0, \ e \sim [A_0, A_1];
\end{aligned} \tag{36}
$$

---

[2] It is possible that scenario (a) may be relevant to the case of normal hierarchy for both flavors $\bar{u}$ and $\bar{d}$, i.e., the formation of CKM matrix for quarks.

where $\alpha_{ik}$ are angles between $A_i$ and $A_k$. According to Eq.(35), the vector $A_2$ is orthogonal to the vectors $A_0$ and $A_1$, $\cos\alpha_{12} = 0$. In Fig.2, axis $Z$ coincides with the direction of $A_2$.

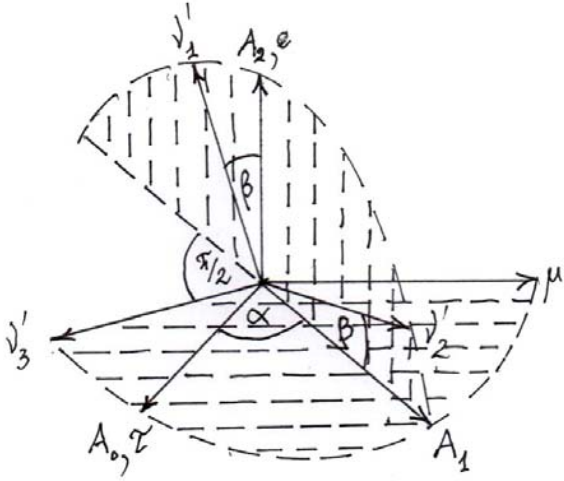

Fig.2: Formation of the "skeleton" of the lepton WMM—the *PMNS* matrix—by rotation between the axes (36) in generation space: $A_2 \perp A_0, A_1$, the angles $\alpha$, $\beta$ are Eqs.(37), (38).

The mixing matrix is the matrix of transition from the $\nu$ basis to the $\ell$ basis (36). Apparently, in the approximation being considered, we have:

$$V = \begin{vmatrix} 1 & 0 & 0 \\ 0 & \cos\alpha_{01} & \sin\alpha_{01} \\ 0 & -\sin\alpha_{01} & \cos\alpha_{01} \end{vmatrix}. \tag{37}$$

Eq.(37) would serve as the basis for the lepton WMM if the neutrino spectrum obeyed a strict hierarchical order (23). The subsequent approximations would lead to small values for those elements that are equal to zero in (37), as well as minor changes for other elements.

For real spectrum (24), one more coordinate rotation, Fig.2, is required to build the "skeleton" of the WMM in the lowest approximation of the hierarchy. One can consider the scenario (24) as the degeneracy of levels "1,2" and proceed with solving a problem of degeneracy removal. This problem was investigated in [2], where we determined conditions required for the parameters (19) to lead to the equality of the masses $m_1 \sim m_2$ in the zero approximation of the hierarchy, as well as the corrections removing degeneracy. The next step is to find correct wavefunctions $\phi_1'$ and $\phi_2'$ again of the zero approximation. The result is well known and consists in the rotation of the degeneracy problem functions $\Phi_1^0$ and $\Phi_2^0$ about some angle $\beta_{12}$ in their plane (Fig.2). In our scenario, these functions coincide with (25); they also do not depend on flavor for the spectrum (24). We have:

$$\phi'_1 = \Phi_1^{(0)} \cos\beta_{12} + \Phi_2^{(0)} \sin\beta_{12} \,,$$
$$\phi'_2 = -\Phi_1^{(0)} \sin\beta_{12} + \Phi_2^{(0)} \cos\beta_{12} \,. \tag{38}$$

The angle $\beta_{12}$ is the function of the parameters of vectors $A$ and $B^{(\nu)}$, so it depends on the flavor $\nu$. However, the actual form of this dependence does not change the general structure of the matrix $V$ and therefore has no importance in our consideration. Upon the rotation (38), the lepton WMM "skeleton" acquires the form:

$$
V = \begin{array}{c|ccc|c}
 & 1 & 2 & 3 & \\
\hline
 & \cos\beta & \sin\beta & 0 & e \\
 & -\sin\beta\cos\alpha & \cos\beta\cos\alpha & \sin\alpha & \mu \\
 & \sin\beta\sin\alpha & -\cos\beta\sin\alpha & \cos\alpha & \tau
\end{array} \tag{39}
$$

Here, we omitted the indices of the angles $\alpha$ and $\beta$. From (39), it is apparent that the element $V_{e3} = |\sin\theta_{13}|$ [1] will be other than zero only when the small terms of the subsequent hierarchical orders are taken into consideration. Estimating its value requires taking into consideration not only the known ratios of charged leptons

$$\frac{m_\ell}{m_\mu} \simeq \frac{1}{207} \simeq 0.0018 \,, \quad \frac{m_\mu}{m_\tau} \simeq \frac{1}{17} \simeq 0.056 \,, \tag{40}$$

but also the corrections for the neutrino wavefunctions (25) and (28) produced by the unknown neutrino mass $m_3 \ll m_1, m_2$. The too small corrections from charged leptons (40) may be insufficient for the description of the observed $|\sin\theta_{13}| \sim 0.14 - 0.16$. The masses $m_1, m_2$ are calculated here from the observed $\Delta m_{ik}^2$ [1]. Then

$$m_3 \approx 0.01 \text{ eV}, \quad \frac{m_3}{m_{1,2}} \lesssim 0.2 \,, \quad m_1 \sim m_2 \sim 0.05 \text{ eV} \quad , \tag{41}$$

appear to be suitable. The remaining elements $V$ can easily be matched with the corresponding values of the *PMNS* matrix. Taking small corrections into consideration will have little impact on other terms of the skeleton (39).

The small value of $m_3 \ll m_1, m_2$ is a necessary condition for the appearance of the structure (39). If this value were not small, the neutrino wavefunctions (25) and (38) would have to include terms defined not only by the $(\nu, \ell)$-flavor independent factor $A$ but also by other factors. The difference of neutrino mass squares [1], the only experimentally known data, would leave such a possibility. This would mean a complete modification of the scenario and the resultant construction of the matrix $V$.

Note that lepton WMM complexities, in our MS scenario, can appear not only from complex Yukawa matrices but also from the Majorana couplings and masses $\mathcal{M}$. Neutrinos formed by two Majorana particles with equal masses are Dirac ones, although, at the same time, they preserve, in principle, the possibility to introduce Majorana phases into the system.

## 5. Conclusions

Discussion in Section 4 allowed us to avoid using almost all factors belonging to the system of neutrino states with the exception of the inverse spectrum order and the hierarchically small neutrino mass $m_3$. This possibility is, of course, a direct result of the chosen MS scenario and its violation. Importantly, we have here the independence from $\bar{u}$-, $\bar{d}$-flavors of the MS-masses $A$ and matrices $U_{LR}$ diagonalizing Yukawa couplings.

The most possible general form of Yukawa couplings would be diagonalized with two unitary matrices

$$h^{(f)} = U_{LR}^+ h_{diag}^{(f)} U_{RL}^{(f)}. \tag{42}$$

Parity conservation in MS models requires that the matrices $h$ (and $\mu$) be Hermitian, and leaves only one flavor-independent unitary matrix in (42):

$$U_{LR} = U_{RL}^{(f)} \equiv U. \tag{43}$$

In the mass terms of MS models, only $B^{(f)}$ masses, of isosinglets, remain dependent on $f$. At the same time, with the Yukawa components of mirror particle masses being different, $\mu_{diag}^{(\ell)} \neq \mu_{diag}^{(\nu)}$, the independence of $U$ from flavor is not easily perceivable for any scenario by which these characteristics appear. It would be more natural to have these parameters equal. Then, use of only charged characteristics in investigations of neutrino systems would be totally understandable.

The second point that is worth discussing here is the chosen form of the matrix (26). The general form of the Hermitian matrix with eigenvalue hierarchy is:

$$h = U^+ h_{diag} U, \quad h_{diag} = (M, m_1, m_2), \quad M \gg m_1 \gg m_2, \tag{44}$$

where $U$ is an arbitrary unitary matrix. In Section 4, we investigated a different, less generalized representation of the Hermitian $h$. In the matrix (44), all elements are simultaneously large, $\sim M$, compared to the form (26). The coefficients of the characteristic equation for (44) are sums of a large number of mutually cancelling contributions; the result of these cancellations being the required form (27).

This situation is similar to the hierarchy obtained from the sum of the so-called "democratic" matrices [13], each of which has one non-zero eigenvalue. These eigenvalues differ in magnitude of hierarchy orders. The matrix (44) is close to the separable forms (8) and (19); all their elements are also large. But the mass hierarchy requires only one large parameter to define the general scale. In any dynamic scheme, the appearance of this parameter in the form of a diagonal element seems to be most acceptable. On the other hand, creating one more level of "mirror symmetry" to explain separability or the "democratic" character of $h$ appears to be incredible.

To solve this problem for neutrino, one would have to deal with simultaneous diagonalization of Yukawa and Majorana interactions. This issue was already noted in Section 3. In MS symmetry, the general forms of the Hermitian and symmetrical matrices (for masses $\mu^{(f)}$ and $\mathcal{M}^{(\nu)}$) are:

$$\mu^{(f)} = U^+ \mu^{(f)}_{diag} U \, ; \quad \mathcal{M}^{(\nu)} = U^+_{\mathcal{M}} \mathcal{M}^{(\nu)}_{diag} U_{\mathcal{M}} + \text{c.c.} \tag{45}$$

The equality

$$U_{\mathcal{M}} = U \tag{46}$$

must be fulfilled, since the two factors (45) together create a single mirror Dirac particle. This is necessary as only the Dirac state leads to Eq.(19), which provides the smallness of neutrino masses and inverse spectrum order.

It should be mentioned that the Dirac nature of mirror and normal neutrinos, corresponding to (19), was proven in [2] only for the lowest approximation of hierarchy. However, next approximations would lead, in high orders, to close in mass, almost degenerate pairs of Majorana particles. The qualitative picture of the scenarios responsible for neutrino properties changes little. This system, however, appears to be less natural. High orders could support the Dirac nature as a result of choosing concrete values of the parameters and not by virtue of the general properties of the model, as it has occurred until now. It is possible that this will allow estimation of numerical values for those parameters.

In conclusion, let us summarize the major phenomenological properties required for reproduction of the lepton WMM structure within the framework of the mirror scenario. They are:

1. The observed mass hierarchy of charged leptons;

2. The Dirac nature of neutrino, predicted by the model, and inverse spectrum of their masses.

Reproduction of the lepton WMM also imposes the following conditions on the MS model:

- Parity conservation in the MS-Lagrangian and the MS violation mechanism (excluding weak interaction);

- Very heavy mirror ($M \gg m_{SM}$) leptons with mass hierarchy that is inverse to the charged lepton hierarchy of SM;

- Involvement of Majorana terms in formation of the neutrino spectrum; non-conservation of the total lepton number for heavy mirror neutrinos [11].

It is difficult to find a dynamic mechanism responsible for the appearance of the poorly pronounced characteristics of the lepton WMM. The spontaneously broken MS scenario may offer an opportunity to find a way to reproduce these properties. Can we expect alternative approaches?

The author is grateful to Ya. I. Azimov for his interest in this work and for useful discussions. This work was funded by grant RSF No. 14-92-0028.

## References

1. Particle Data Group, Chin. Phys. C **40**, 100001 (2016).

2. I. T. Dyatlov, Yad.Fiz.**78**, 522 (2015) [Phys. Atom. Nucl. **78**, 485 (2015)]; arXiv:1502.01501[hep-ph].

3. I. T. Dyatlov, Yad.Fiz.**78**, 1015 (2015) [Phys. Atom. Nucl. **78**, 956 (2015)]; arXiv:1509.07280[hep-ph].

4. T.D. Lee and C. N. Yang, Phys. Rev. **104**, 254 (1956).
   J. Maalampi and M. Roos, Phys. Rept. **186**, 53 (1990).
   L.B. Okun, UFN **177**, 397 (2007) [Phys. Usp. **50**, 380 (2007)] hep-ph/0606202.
   P.Q. Hung, Phys. Lett. B **649**, 275 (2007)
   Pei-Hong Gu, Phys. Lett. B **713**, 485 (2012)
   S. Chakdar *et al.*, arXiv:1305.2641 [hep-ph].

5. I. T. Dyatlov, Yad.Fiz. **77**, 775 (2014) [Phys. Atom. Nucl. **77**, 733 (2014)]; arXiv:1312, 4339[hep-ph].

6. R. N. Mohapatra and A. Y. Smirnov, hep-ph/0603118;
   S. F. King *et al.*, arXiv:1402.4271 [hep-ph].
   L. Maiani, arXiv:1406.5503[hep-ph].

7. C.D. Froggatt, M. Gibson, H.B. Nielsen and D.J. Smith, hep-ph/9706212;
   C.D. Froggatt and H.B. Nielsen, hep-ph/9905445;
   D. I. Silva-Marcos, hep-ph/0102079;
   F. Sannino, arXiv:1010.3461[hep-ph];
   G. C. Branco *et al.*, arXiv:1101.5808[hep-ph];
   Zhi-zhog Xing, arXiv:1411.2713[hep-ph].

8. F. P. An *et al.* [Daja Bay Collab.], Phys. Rev. Lett. **108**, 171803 (2012);
   J. K. Ahn *et al.* [RENO Collab.], Phys. Rev. Lett. **108**, 191802 (2012);
   Y. Abe *et al.* [Double Chooz Collab.], Phys. Rev. D**86**, 052008 (2012)

9. K. Abe *et al.*, [T2K Collab.], Phys. Rev. D **91**, 072010 (2015).

10. G. Altarelli, arXiv:1404.3859 [hep-ph];
    Shun-Zhou, arXiv:1511.07255 [hep-ph].

11. I. T. Dyatlov, Yad.Fiz. 80, No.1 (2017); arXiv:1611.05635 [hep-ph].

12. L. Wolfenstein, Phys. Rev. Lett. **51**, 1945 (1983).

13. H. Fritzsch, in: Proceed. Of the Europhys. Topical Conf. on Flavour Mixing in Weak Interactions, Erice, Italy, 1984, Ed. by L. L. Chaw (Ettore Majorana Intern. Series, Phys. Science, Vol. 20; Plenum Press, NY, 1984).